\documentclass[10pt]{iopart}
\usepackage{iopams} 

\usepackage{graphics}
\usepackage{graphicx}
\usepackage{color,soul}
\begin{document}

\title{Muon Detector for Underground Tomography}

\author{
Yan  Benhammou$^{1}$,  Erez  Etzion$^{1}$, Gilad  Mizrachi$^{1,2}$, Meny Raviv Moshe$^{1}$, Yiftah  Silver$^{2}$ and Igor Zolkin$^{1}$}

\address{$^{1}$ Raymond and Beverly Sackler School of Physics \& Astronomy, Tel Aviv University, Tel Aviv 69978, Israel,\\ $^{2}$ Rafael, Advanced Defense systems LTD, Haifa, Israel}

\ead{erez@cern.ch}
\ead{giladmiz01@gmail.com}

\vspace{10pt}
\begin{indented}
\item[]May 2022
\end{indented}

\begin{abstract}
We utilise muons from cosmic ray  to explore hidden underground   archaeological structures. Presented here is the design, simulation studies and first laboratory results of a compact, scintillators based, cosmic ray muon telescope for underground muon radiography.  

\end{abstract}

%
\vspace{2pc}
\noindent{\it Keywords}: High Energy Physics, Particle Detectors, Muon Tomography, 3-D imaging


\ioptwocol
\section{Introduction}
\label{sec:Introduction}
The utilisation of cosmic ray muons for radiography and tomography  was first applied in the 1950s by E. P. George to measure
the overburden over a tunnel in Australia~\cite{tunnelTomography:1}, and in the late 1960s' by Luis Alvarez in his famous attempt to
discover hidden chambers in the second pyramid of Chephren in Giza~\cite{Alvarez832}.
Recent advances in detector technology and data processing have brought the concept of muon tomography to
a point where it can be used reliably for subsurface imaging of geological or mechanical structures~\cite{Schouten:2018kla}. This was recently demonstrated in
several volcano imaging campaigns, deposits explorations and archaeological surveys,  among which are:~\cite{Schouten:2018pqz,Morishima:2017ghw, MURAVES:2020lgx,tanaka:06}
that provided subsurface density profiles. The detected variations are of the order of a few percent in density contrast, making it a powerful tool for discriminating between rocks of different densities, voids, water or other substances.
Cosmic ray muon subsurface imaging can be done with many different technologies  all of which rely on several factors:

\begin{itemize}
\item Nearly constant flux of muons reaching the surface and well known energy spectrum and angular distribution (e.g \cite{Reyna2006ASP}).
\item Muons traverse through the subsurface material approximately along straight lines, so that high (enough) angular resolution of the $\mathcal{O}$(25 mrad) can be reached~\cite{Zyla:2020zbs-1}. 

\item Muons lose energy at nearly constant pace when they penetrate through the Earth’s surface. This results in a well-known intensity of muons as a function of the angular direction and the effective depth~\cite{Zyla:2020zbs-2}.   

\end{itemize}

Muon imaging relies on the known correlation between  the measured muon flux and the density length of the traversed material. The flux measured at each solid angle is compared to the predicted flux (the one that would have been observed with homogeneous ground) and anomalies in the subsurface modify the observed rate. By the use of several detectors at different locations, a 3-D image (tomography) is achieved. 

While the focus of the current work is a compact detector suited for muon tomography in archaeological sites, it serves as a small prototype for the much larger scale, MATHUSLA experiment~\cite{lubatti2019mathusla,MATHUSLA:2020uve,MATHUSLA:2022sze,ALIDRA2021164661}, aiming at detection unknown particles escaping detection at the LHC experiment at CERN.

\section{Detector design}

Muon detection can be achieved in two steps: In the first, using scintillation, the charged particle (i.e. muon) produces photons along its' path in the scintillating material. In the second step, the emitted light is being collected (via Wave Length Shifting, WLS, fiber) and transformed into measurable electronic signal with a light detector such as Silicon Photomultiplier (SiPM). 

The detector we present here is constructed of four layers of scintillating bars: Two perpendicular layers at the top of the detector and two perpendicular layers at the bottom of the detector. When a muon passes through a layer it deposits enough energy (i.e. enough photons are registered at the SiPM) to estimate the hitting point with $\mathcal{O}$(0.5 cm) resolution (the procedure will be explained in the following sections). As a consequence, when a muon passes through all four layers it is possible to reconstruct it's trajectory with an angular resolution sufficient for successful underground muon radiography. 

\subsection{Basic Building Block - Extruded scintillator, WLS fiber and SiPM}
\begin{figure}[ht!]
\begin{center}
 a) \includegraphics[height=5cm]{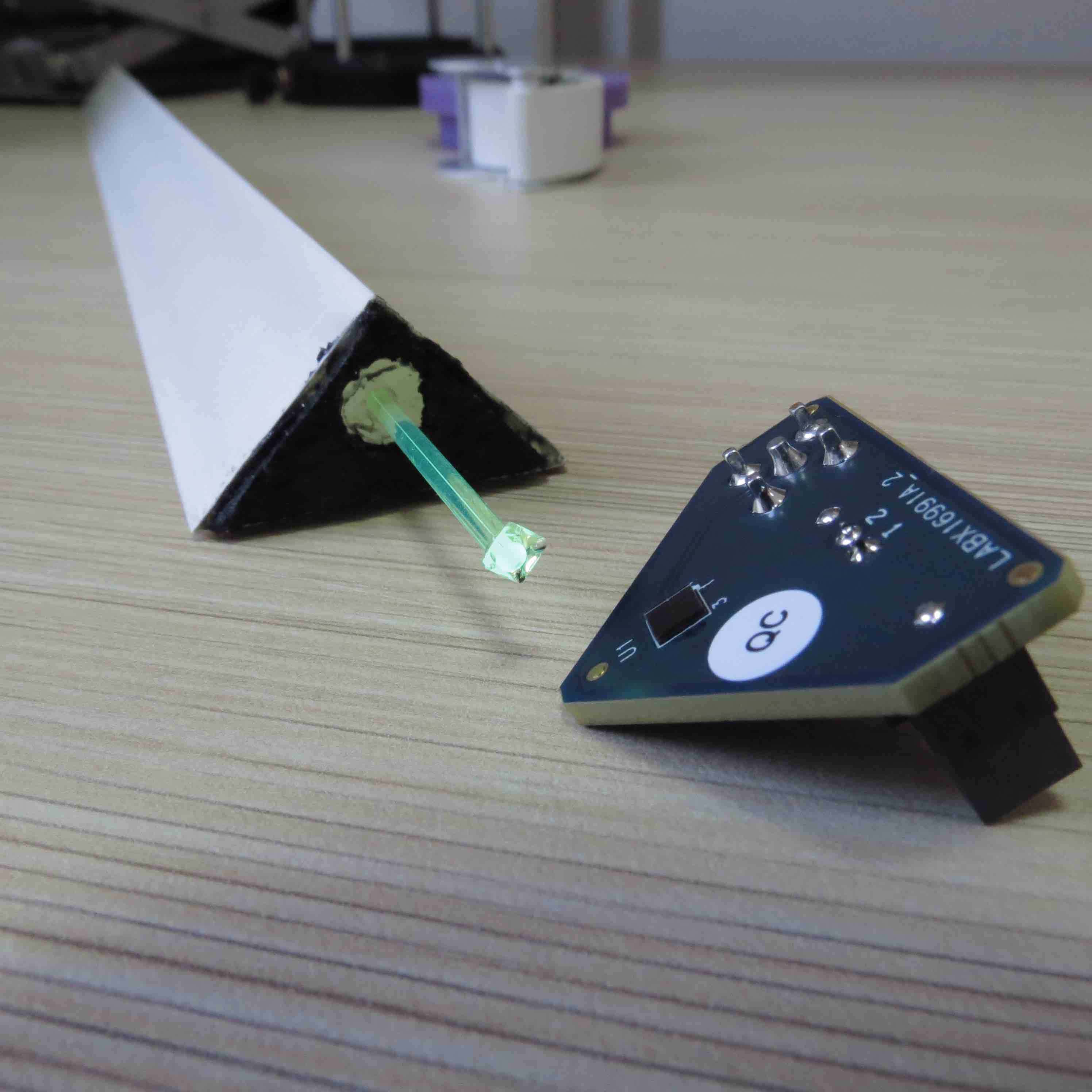}
 b) \includegraphics[height=5cm]{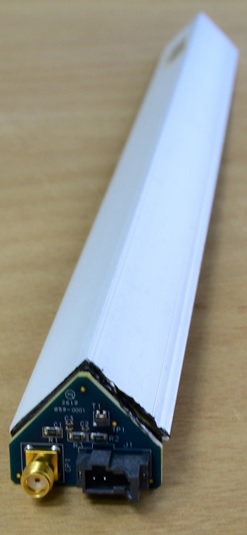}
\end{center}
\caption{
 Triangular extruded scintillator with WLS and a SiPM before (a) and after (b) the attachment to the Front-end board.
}
\label{fig.bar}
\end{figure}

A single "pixel" is comprised of a 3.3~cm $\times$ 1.7~cm $\times$ 40~cm triangular bar of extruded plastic scintillator made out of general purpose Polystyrene, doped with 1\% PPO and
0.03\% POPOP, and coated with titanium-oxide~\cite{Pla-Dalmau:2000puk}. At some point we have considered to utilise liquid scintillating detectors but realised that extruded plastic bars are more convenient to handle than our attempt of using bars made of liquid scintillating containers~\cite{Benhammou_2020}.  The scintillating light is collected by WLS optical fiber (Saint-Gobain BCF-91A)  inserted into a central hole in the bar and attached (using Saint-Gobain BC-631  optical grease) to the face of a SiPM (Sensel microJ 30035) \cite{Onsemi}. 
The SiPM is mounted on a triangle-shaped printed circuit board (PCB) that also contains the electrical connectors for power supply and the SiPM analog output, see figure \ref{fig.bar}. 

\begin{figure}[ht!]
\begin{center}
a) \includegraphics[width=0.86\linewidth]{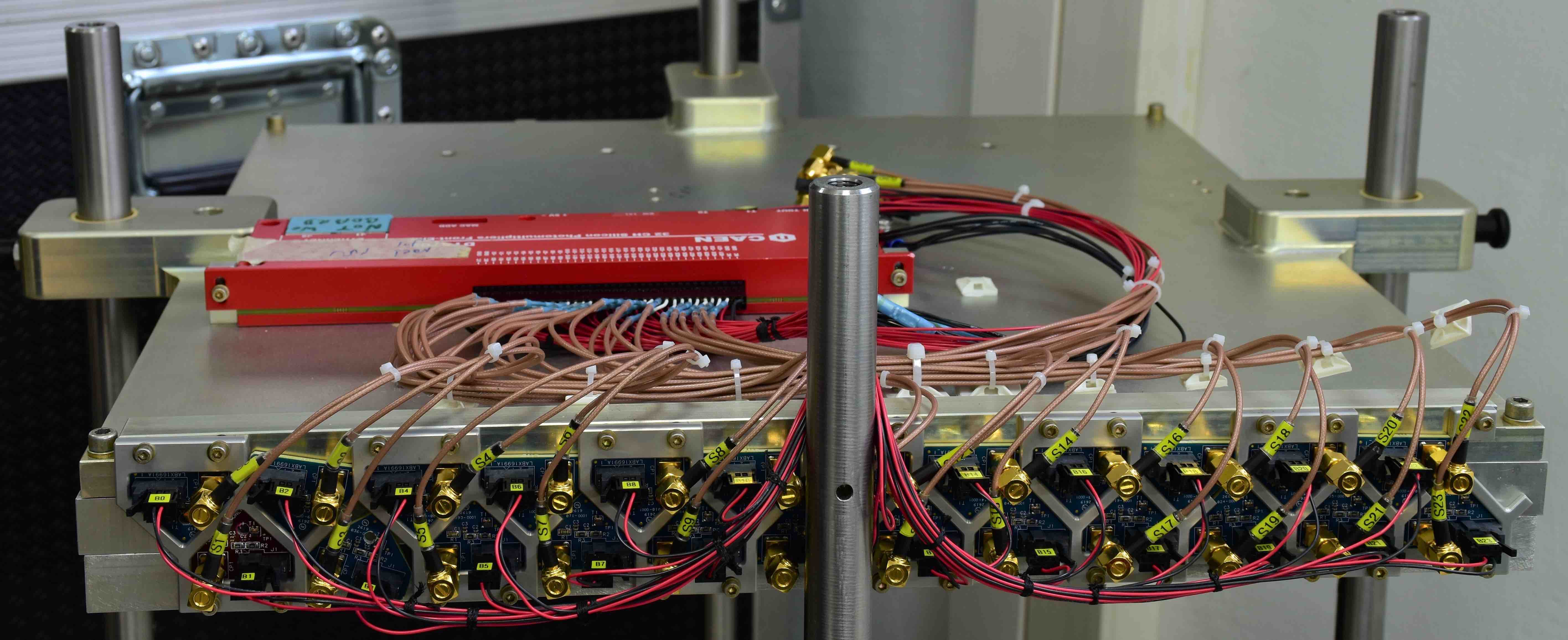}

b) \includegraphics[width=0.86\linewidth]{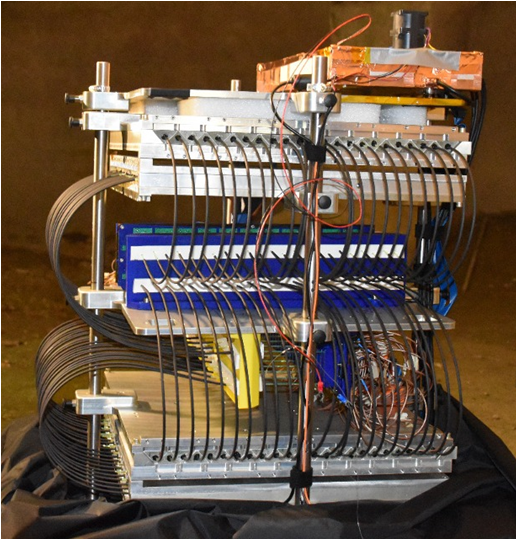}
\end{center}
\caption{
 (a) The first version of a single layer of the telescope, consisting of bars, mechanical enclosure and the front-end electronics; (b) Picture of the complete detector (four layers) in the final design.
}
\label{fig.layer}
\end{figure}

The triangular PCB was originally designed to be attached directly on the scintillator as shown in figure~\ref{fig.bar}. However, we have changed our design due to the following: In general there are two typical attenuation lengths for the WLS fibers; the long one (which is generally cited) is of $\mathcal{O}$ (meters) and a short attenuation length of $\mathcal{O}$ (10s cm)~\cite{David:1994rwa}. The presence of the short attenuation length demands a very broad dynamic range for the data acquisition. Therefore, we have modified our original design to facilitate a simple data acquisition (DAQ) readout system with moderate dynamic range on the expanse of longer fibers. Placing about half a meter of WLS fiber between the scintillator and the SiPM allows for the effects of the short attenuation length to be diminished sufficiently. Pictures of the original design and the final one are shown in figure~\ref{fig.layer}.

\subsection{Data Acquisition}
Power supply for the SiPMs as well as the signal read out is done using CAEN's DT5550W DAQ platform~\cite{CAEN}.  

The DT5550W is a desktop programmable complete readout system, based on WeeROC ASICs composed of: a mezzanine card that hosts four 32 channels Citiroc ASICs, programmable Xilinx XC7K160T FPGA designed to read out the ASICs, power supply and ADCs, on board 128 lines of power supply for the SiPM Bias (20-85 V).

The Chosen platform has an on board 14-bit 80 MS/s simultaneous sampling ADC, to monitor and acquire the analog outputs from each SiPM (i.e. energy and time measurements). 

\begin{figure}[h!t]
\begin{center}
\includegraphics[width=0.85\linewidth]{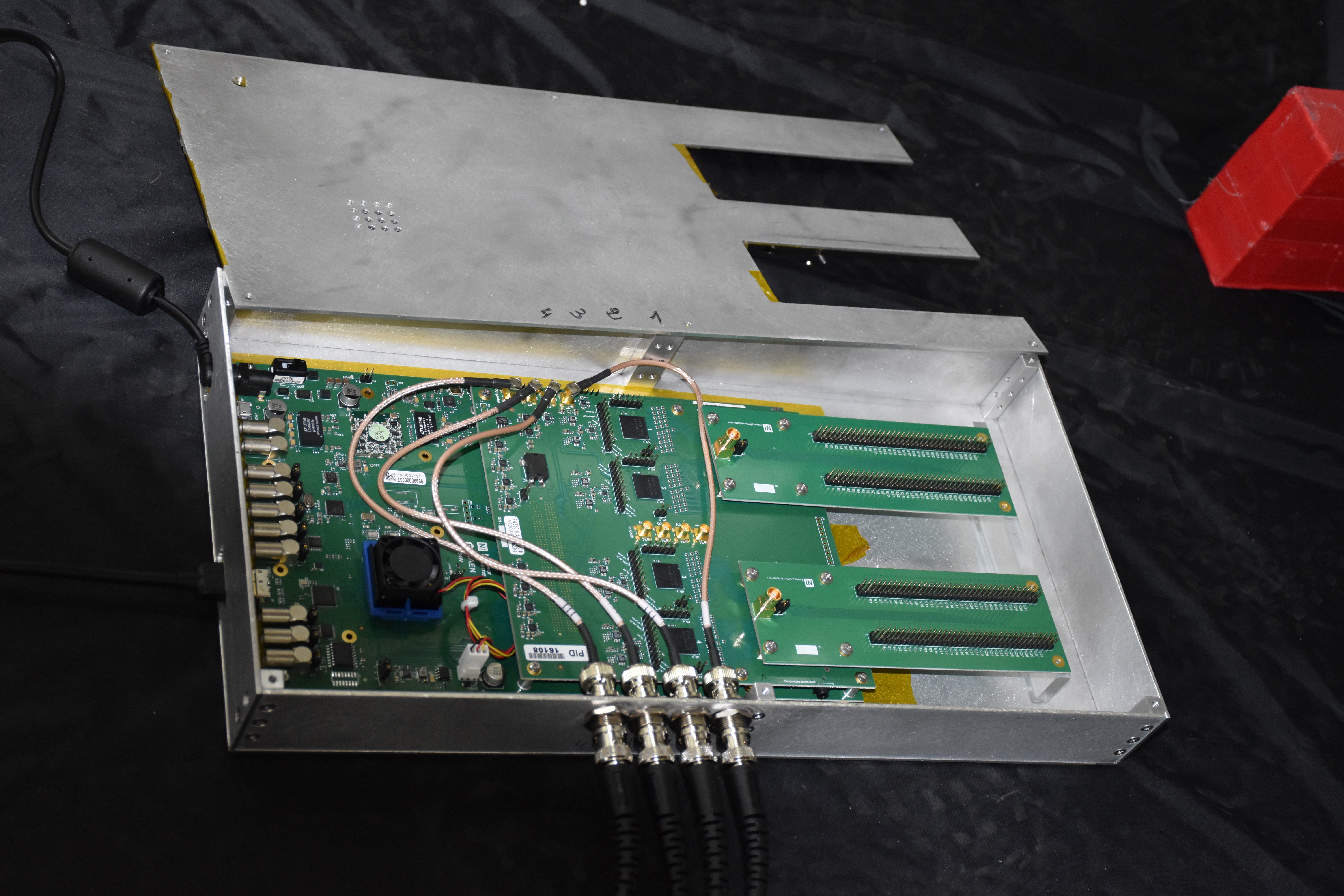}
\end{center}
\caption{A picture of the electronic DAQ, CAEN DT5550W in a dedicated heat dissipation box.
}
\label{fig.DAQBOX}
\end{figure}

The described DAQ is mounted in a dedicated box tailored especially for efficient heat dissipation; The DAQ is shown in figure~\ref{fig.DAQBOX}.

\subsection{Enclosure and detector construction}
A single detector layer comprised of 23 triangular bars arranged as shown in figure~\ref{fig.layer}. Each layer is encapsulated in an Aluminum frame, for mechanical strength and for light insulation. All four detector layers are mounted in a steel frame stand enabling a precise vertical positioning (z axis) of each layer. Upon installation the complete detector is instrumented inside a light tight box of 65~$\times$~65~cm$^2$ and height of 80~cm.

\section{Performance}
\subsection{Muon reconstruction}
\begin{figure}[htb]
    \begin{center}
        \includegraphics[width=0.8\linewidth]{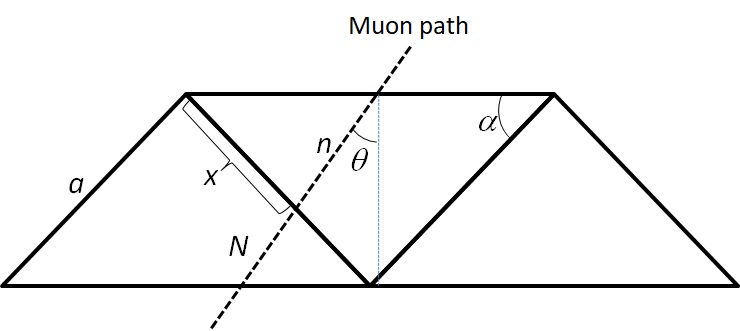}
    \end{center}
    \caption{
     Muon passes through two neighbor bars creates two similar triangles. Knowing $n$, $N$, and $a$ allow us to calculate $x$.
    }
    \label{fig.TriangleBarExplanation}
\end{figure}

The number of emitted photons in the scintillator is proportional to the path length of the muon through the scintillator. 
As the muon passes through the layer it generates a signal in one or more bars. This allows to determine one coordinate of the hitting point of the muon in the layer. Furthermore, by using triangular shaped scintillators arranged  as seen in figure ~\ref{fig.TriangleBarExplanation} one can achieve sub cm resolution of the hitting point \cite{MINERvA:2013zvz}.

A muon which enters the bar in an angle {$\theta$} with respect to the zenith, and crosses the interface between the sides of two bars at a distance $x$ from the triangle vertex (see figure ~\ref{fig.TriangleBarExplanation})  generates $n$ photons at the first bar and  $N$ photons at the second bar. $N$ and $n$ are proportional to the path length of the muon in each bar. It is easy to see that the muon path creates two similar triangles and hence we can write the following:
\begin{equation}
    \frac{n}{N} = \frac{x}{a-x}
    \label{eq:SubPixelEq}
\end{equation}
where  $a$ is the triangle side length. We can extract $x$ and get a better estimation of the hitting point.

\subsection*{Error estimation for the hit position}

By using equation~\ref{eq:SubPixelEq} we determine the  hitting coordinate of the muon in the layer. However, as $n$ and $N$ are of Poissonian nature  the uncertainty in  $x$  is:
\begin{equation}
    \Delta x^2=\left(\frac{\partial x}{\partial N}\cdot \Delta N\right)^2+\left(\frac{\partial x}{\partial n}\cdot \Delta n\right)^2
\end{equation}
which translates (assuming $\Delta N=\sqrt{N}$ and $\Delta n=\sqrt{n}$) to:
\begin{equation}
    \label{eq:SubPixelErrorEq}
    \Delta x=\frac{a}{N+n}\sqrt{\frac{n \cdot N}{N+n}}.
\end{equation}

\begin{figure}[!htb]
    \begin{center}
        \includegraphics[width=8cm]{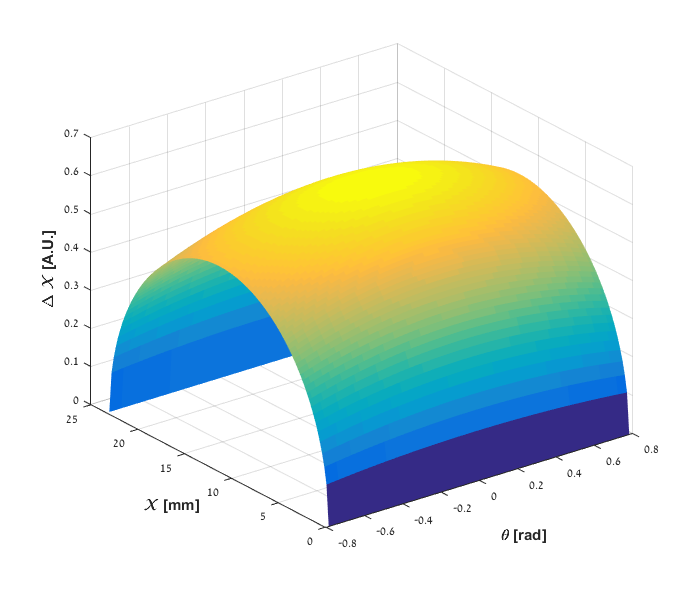}
    \end{center}
    \caption{
     Expected  spatial and angular uncertainty.
    }
    \label{fig.ExpectedDeltaX}
\end{figure}

In order to evaluate $\Delta x$, we need to measure the number of photons emitted along the scintillator while a muon crosses it. The expected behaviour of $\Delta x$ and its dependency on $x$ and $\theta$, is demonstrated in figure~\ref{fig.ExpectedDeltaX}.

\begin{figure}[!htb]
    \begin{center}
        \includegraphics[width=8cm]{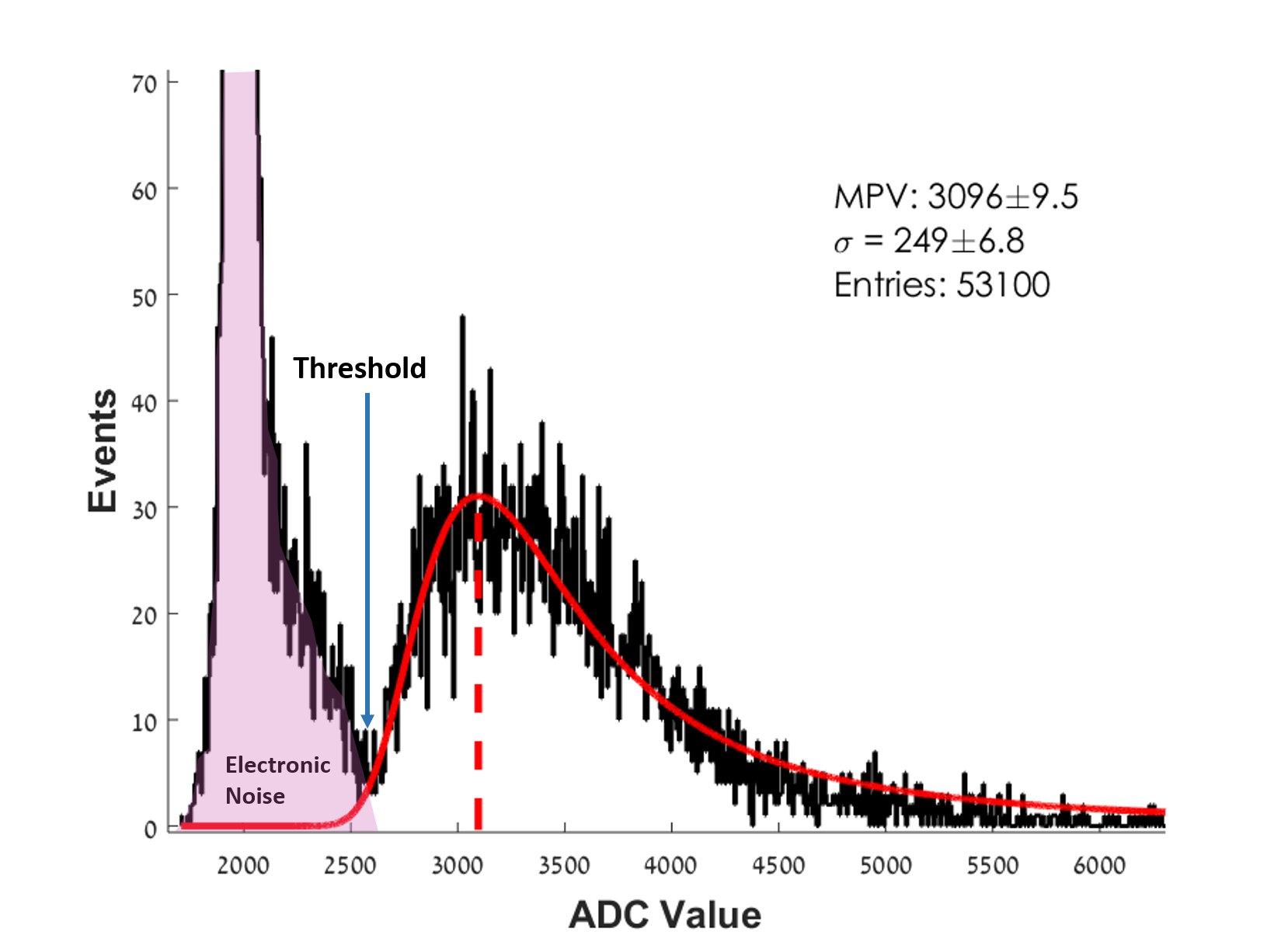}
    \end{center}
    \caption{
     Typical raw data (ADC counts) recorded for one triangular bar. the electronic noise area is coloured. A Landau fit is performed on the accumulated signals, and it's Most Probable Value (MPV) is used for gain calibration and tuning (done independently for every channel)  
    }
    \label{fig.threshold}
\end{figure}

To measure the hit resolution for each detector layer and validate the above estimation we have done the following:

\begin{enumerate}
    \item Four detector layers were assembled at the same orientation, i.e. all four layers are parallel to the $\hat{x}$ axis (unlike to the normal detector assembly in which two layers are perpendicular to the others).
    \item $\mathcal{O}(10^6)$ events were recorded; An event consists of the raw signal registered in all the bars. At least one  bar with signal above the threshold is required to record an event. The threshold is set above the electronic noise as seen in figure~\ref{fig.threshold}.   
    \item $\mathcal{O}(10^5)$ events pass a selection criteria of at least one bar hit (and no more than two adjacent bar hits) for all four layers.
    
    \item Hit position is estimated using equation \ref{eq:SubPixelEq} for every layer, resulting with a set of $(x_i,z_i)$. 
    
    \item A straight line fit is performed for three out of the four layers, and the residual is calculated between the estimated hit position for the remaining layer and the straight line fitting prediction.
    
    \item The hit resolution and any systematic errors are estimated using the residual distributions for each detector layer. 
    
\end{enumerate}

\begin{figure}[htb]
   \begin{center}

        a) \includegraphics[width=0.85\linewidth]{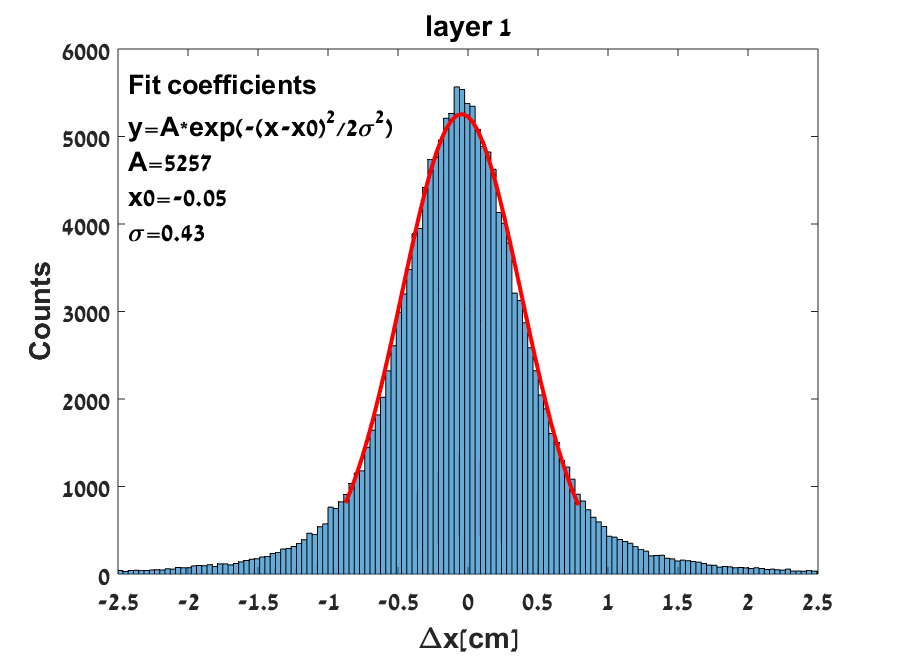}
        
        b) \includegraphics[width=0.85\linewidth]{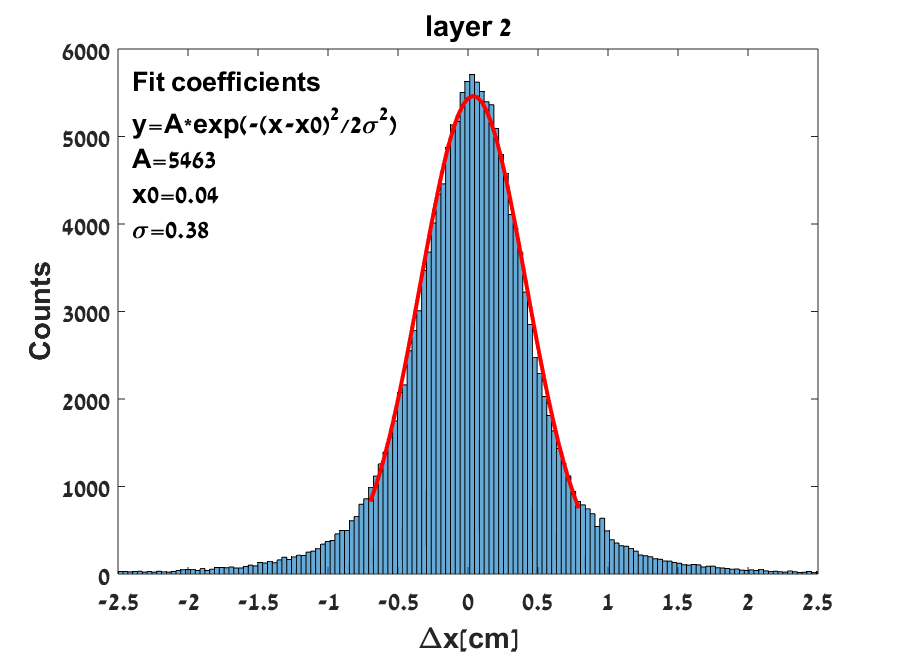}
        
        c) \includegraphics[width=0.85\linewidth]{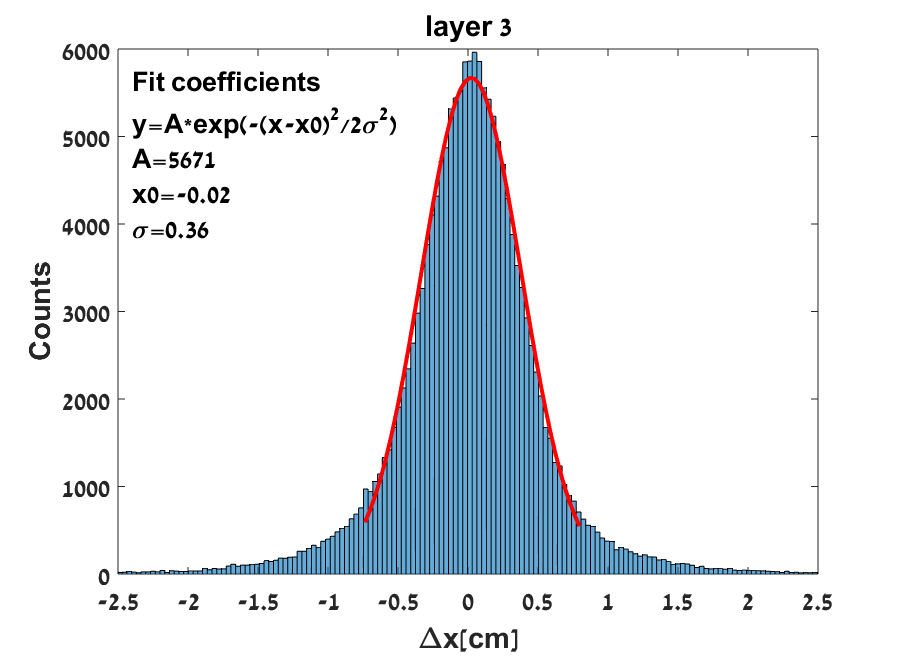}
        
        d) \includegraphics[width=0.85\linewidth]{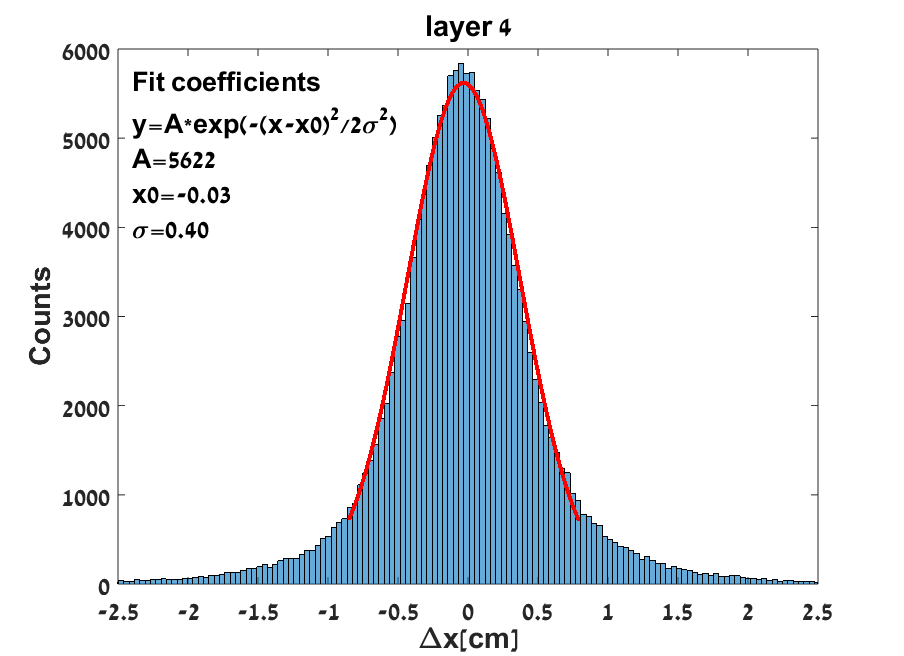}
   \end{center}      
    \caption{
    Residual distributions for each detector layer (a)-(d) and the resulting hit resolution.
    }
    \label{fig.resolution}
\end{figure}

Figure~\ref{fig.resolution} shows the resulting hit resolution for all four layers; As seen in the legends of the figures (a)-(d) the measured resolution ($1\sigma$) is between 3.5 mm and 4.5 mm for all layers.   

\subsection{Simulation studies}
\begin{figure}[!htb]
    \begin{center}
        \includegraphics[width=0.9\linewidth]{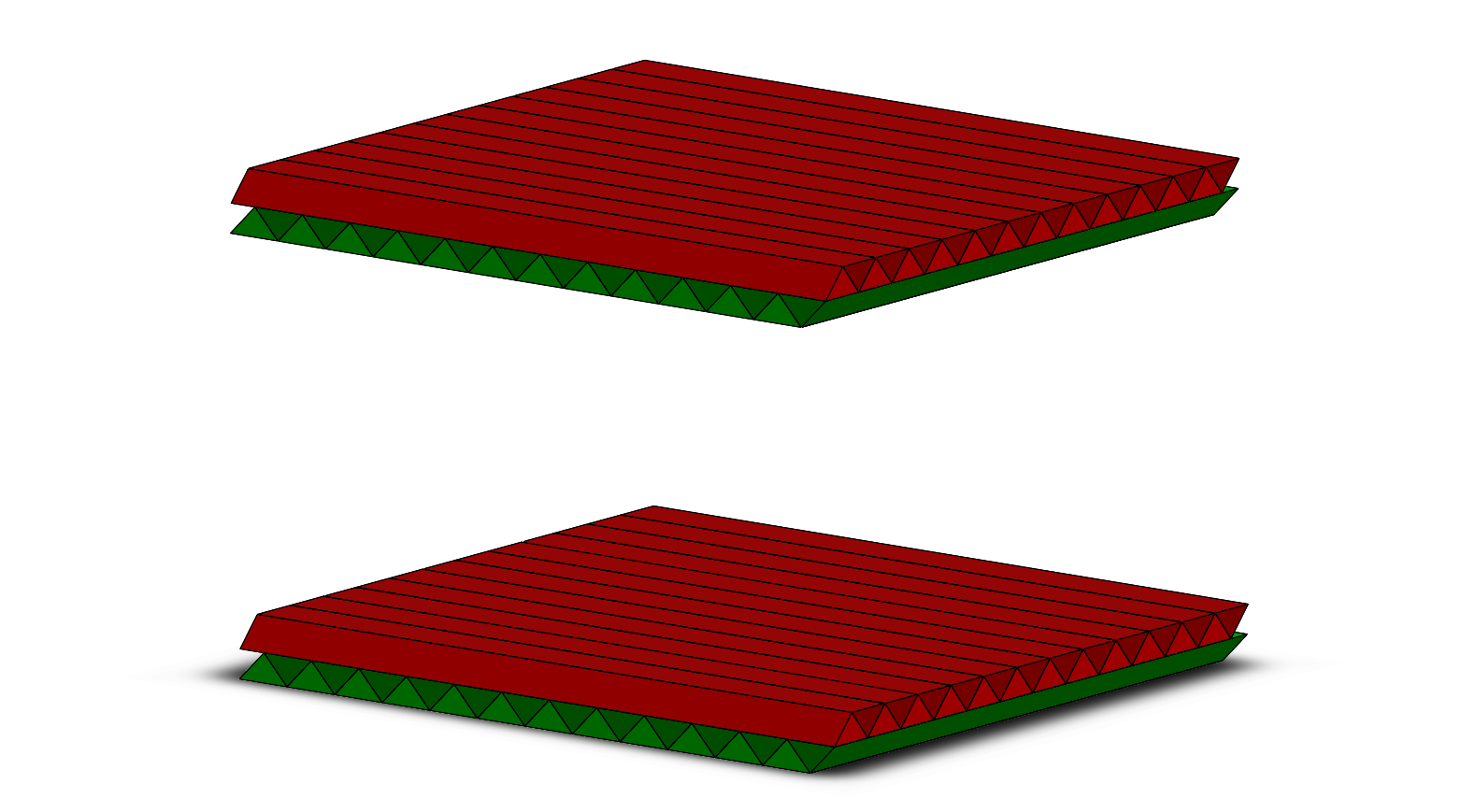}
    \end{center}
    \caption{
    The detector in the GEANT4 simulation, is made of four layers, each of which is made of 23 bars. 
    }
    \label{fig.DetectorAtSimulation}
\end{figure}

\begin{figure}[!htb]
    \begin{center}
        a) \includegraphics[width=0.35\linewidth]{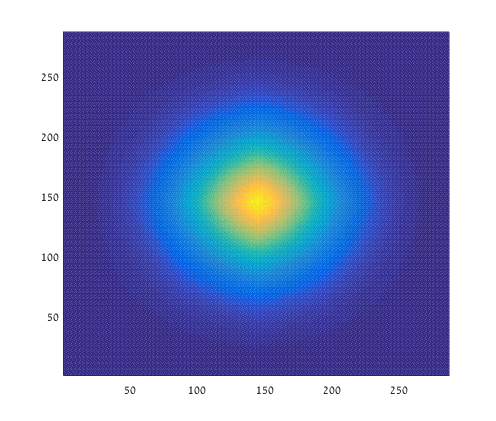}
        b) \includegraphics[width=0.35\linewidth]{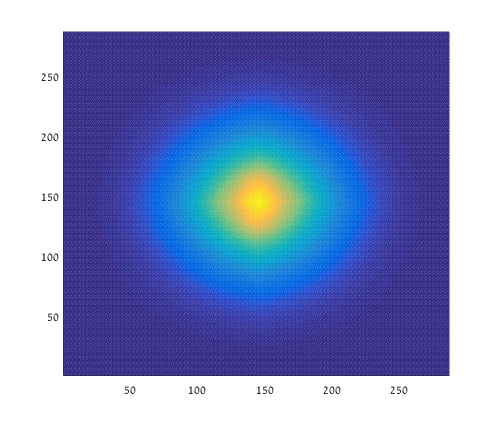}
        
        c) \includegraphics[width=0.5\linewidth]{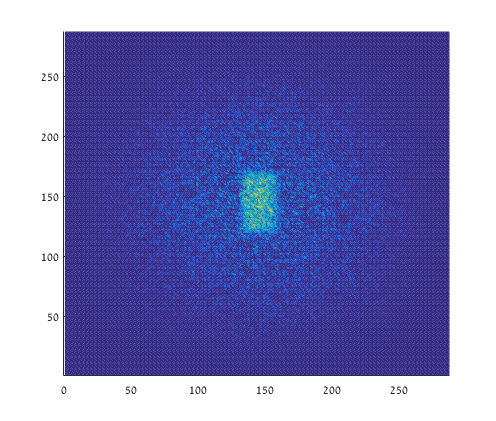}
    \end{center}
    \caption{
     (a) Integration time of approximately two weeks  of a reference model (standard rock, no air filled box), (b) Integration time of approximately two weeks  of a target model (the insertion of an air filled box), (c) subtraction of the two histograms on the left side.
    }
    \label{fig.SimulationResult}
\end{figure}

The detector performance and the tracking algorithm were evaluated using a simulated environment.   We have constructed a complete detector simulation using the GEANT4~\cite{Agostinelli:2002hh} simulation toolkit. The simulated detector is made of four layers, where each layer is made of 23 bars as shown in figure ~\ref{fig.DetectorAtSimulation}. Simulated muons  generated according to the known distributions ~\cite{Reyna2006ASP} pass through the simulated scenarios (ground made of realistic materials containing some known mechanical structures and / or voids) and then through the detector. For each bar the registered signal is proportional to the path length of the muon through the bar. We show here the result of one study where a $2 \times 1 \times 1~ m^3$ air void is located 4~m above the detector. The surrounding volume is made of homogeneous standard rock. 
 Figure~\ref{fig.SimulationResult} depicts the simulation results: (a) Integration time of approximately two weeks  of a reference model (standard rock, without voids), (b) Integration time of approximately two weeks  of a target model (with the air void specified above), (c) subtraction of the two simulated models (a-b) where the air void is clearly seen.

\subsection{Blue sky measurement and comparison to the simulation}
\begin{figure}[!htb]
    \begin{center}
        \includegraphics[width=1.1\linewidth]{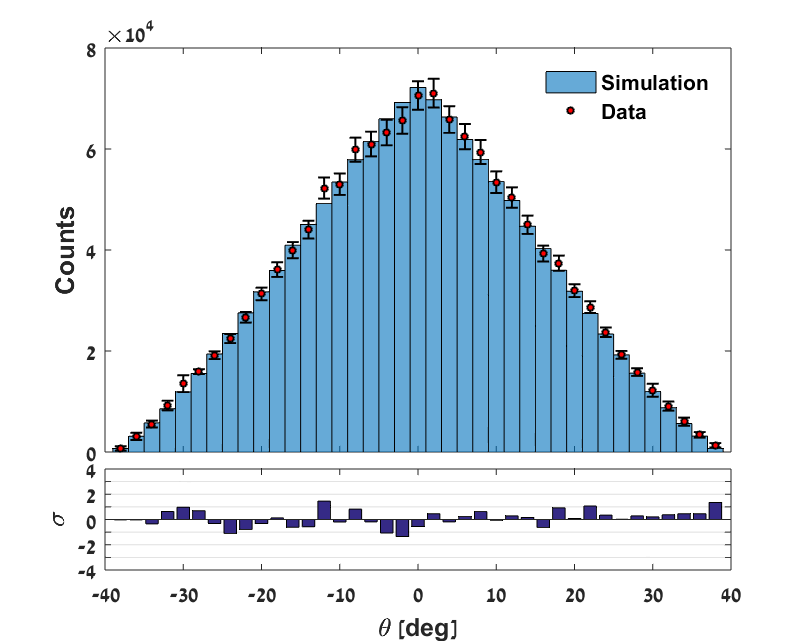}
    \end{center}
    \caption{
    Comparison between 1-D measured muon flux to the simulated one for a blue sky scenario. The bottom shows the tension ($\sigma$) between the data and the simulation.  
    }
    \label{fig.BlueSky}
\end{figure}

In order to validate our detector and analysis procedure we have compared the measured muon flux to the simulated one for a blue sky scenario (i.e. the detector is placed at sea level with minimal overburden). The data depicted in figure~\ref{fig.BlueSky} were taken using the same configuration used in the hit resolution measurements (i.e. four parallel layers) resulting in a 1-D measurement of the muonic angular flux.  Figure~\ref{fig.BlueSky} shows the agreement between the blue sky 1-D data and the GEANT4 simulation. The uncertainties presented in the figure include both the statistical uncertainty ($\sqrt{N}$ of the events in each angular bin) and the systematic uncertainty combined in quadrature.
The systematic uncertainty was found to be $4.3\% \pm 2.0 \%$  and was estimated in the following procedure:
\begin{enumerate}

    \item Re-analyse the raw data from the top and bottom layers; By adding or subtracting a one sigma distance ($\approx 0.4 $ cm) we  generate two new hit position data-sets (modified hit position), for each detector layer.
    \item Re-calculate the angular flux ($\Phi$) using the modified hit positions with positive one sigma adjustment for the top layer, and negative one sigma for the bottom layer in quadrature ($\Phi_{\sigma^+} $).
    \item Re-calculate the angular flux using the modified hit positions with negative one sigma adjustment for the top layer, and positive one sigma for the bottom layer in quadrature ($\Phi_{\sigma^-} $).
    \item Estimating the uncertainty  as $\frac{ | \Phi_{\sigma^+} - \Phi_{\sigma^-} |}{2 \dot \Phi} $  for each angular bin.

\end{enumerate}

\section{Summary and Outlook}
\label{sec:Summary}

A compact and easy to operate muon telescope was tested and validated against comprehensive GEANT4 simulation and was found suitable for underground imaging applications. Specifically, the presented detector is made out of four layers each of which has a hit resolution of approximately 4 mm; The mechanical structure has adjustable separation distance between the different layers enabling angular resolution of approximately 15 - 40 mrad (depending on the vertical distance between the layers). This uncertainty is comparable to the Coulomb multiple scattering deflection angle of  muons penetrating through the ground~\cite{Zyla:2020zbs-1} and reaching  archaeology relevant depths of order tens of meters.
In the next few months we intend to position and commission the detector in the City of David excavation area in Jerusalem, looking for undiscovered voids, conduits or water tunnels (with the guidance of the archaeological team).

\section*{Acknowledgment}
The authors would like to thank Oded Lipschitz and Yuval Gadot from Tel Aviv University Archaeology and Ancient Near Cultures department and  Yiftah Shalev from Israel Antiquities Authority, for the Archaeological justification and motivation as well as the priceless history lessons. We thank Amir Weissbein from Rafael Physics Department and Liron Barak from Tel Aviv University School of Physics and Astronomy for the valuable physics advises and algorithmic assistance.    
The authors would like to thank the PAZY, Kantor and Cogito foundations.

\section*{References}

\bibliographystyle{ieeetr}
\bibliography{Muon-detector-for-underground-tomography.bib}
\end{document}